\documentclass[12pt]{iopart}
\usepackage[dvips]{graphicx}
\usepackage{url}
\usepackage{amssymb}
\usepackage{hyperref}
\begin{document}

\title[Modeling scaled processes and $1/f^\beta$ noise]{Modeling scaled processes and $1/f^\beta$ noise by the nonlinear stochastic differential equations}

\author{B Kaulakys and M Alaburda}

\address{Institute of Theoretical Physics and Astronomy of Vilnius University, Go\v stauto 12, LT-01108 Vilnius, Lithuania}
\ead{kaulakys@itpa.lt}

\begin{abstract}
We present and analyze stochastic nonlinear differential equations generating signals with the power-law distributions of the signal intensity, $1/f^\beta$ noise, power-law autocorrelations and second order structural (height-height correlation) functions. Analytical expressions for such characteristics are derived and the comparison with numerical calculations is presented. The numerical calculations reveal links between the proposed model and models where signals consist of bursts characterized by the power-law distributions of burst size, burst duration and the inter-burst time, as in a case of avalanches in self-organized critical (SOC) models and the extreme event return times in long-term memory processes. The presented approach may be useful for modeling the long-range scaled processes exhibiting $1/f$ noise and power-law distributions. 
\end{abstract}

\maketitle 

\section{Introduction}
The inverse power-law distributions, autocorrelations and spectra of the signals, including $1/f$ noise (also known as $1/f$ fluctuations, flicker noise and pink noise), as well as scaling behavior in general, are ubiquitous in physics and in many other fields, counting natural phenomena, spatial repartition of faults in geology, human activities such as traffic in computer networks and financial markets. This subject is a hot research topic for many decades (see, e.g., \cite{Johnson,Mandelbrot,Mandelbrot2,Voss,Reviews,Wong,Musha,Kaulakys,Kaulakys1,Kaulakys2} and references herein). An up-to-date bibliographic list on $1/f$ noise of more than 1300 papers is composed by Wentian Li \cite{Li}. 

Widespread occurrence of signals exhibiting such a behavior suggests that a generic, at least mathematical explanation of the power-law distributions might exist. Note that the origins of two popular noises, i.e., the white noise -- no correlation in time, the power spectrum $S(f)\sim f^0$, and integral of the white noise -- the Brownian noise (Wiener process), no correlation between increments, the power spectrum $S(f)\sim f^{-2}$, are very well known and understood. $1/f^\beta$ noise with $0<\beta<2$, however, cannot be realized and explained in a similar manner and, therefore, no generally recognized explanation of the ubiquity of $1/f$ noise is still proposed. 

Despite numerous models and theories proposed since its discovery more than 80 years ago \cite{Johnson}, the intrinsic origin of $1/f$ noise still remains an open question. Although in recent years it is annually published about 100 papers with the phrases `$1/f$ noise', `$1/f$ fluctuations' or `flicker noise' in the title, there is no conventional picture of the phenomenon and the mechanisms leading to $1/f$ fluctuations are not often clear. Most of the models and theories have restricted validity because of the assumptions specific to the problem under consideration. Categorization and summary of the contemporary stage of theories and models of $1/f$ noise are rather problematic: on one hand, due to the abundance and variety of the proposed approaches, and on the other hand, for the absence of the recent comprehensive review of the wide-ranging "problem of $1/f$ noise" and because of the lack of a survey summarizing the current theories and models of $1/f$ noise. We can cite only a pedagogical review of $1/f$ noise subject by Milotti \cite{Milotti}. Wentian Li presents some kind of classification by the categories of publications related with 1/f noise until 2007 \cite{Li_Old}. In the peer-reviewed encyclopedia Scholarpedia, \cite{Scholarpedia} there is also a short current review on the subject under the consideration. Thus, we present here only a short and partial categorization of $1/f$ noise models with the restricted list of references. 

Until recently, probably the most general and common models, theories and explanations of $1/f$ noise have been based on some formal mathematical description such as fractional Brownian motion, half-integral of the white noise or some algorithms for generation of signals with the scaled properties \cite{Mandelbrot3} and popular modeling of $1/f$ noise as the superposition of independent elementary processes with the Lorentzian spectra and proper distribution of relaxation times, e.g., $1/\tau_{relax}$ distribution \cite{Bernamont}. The weakness of the later approach is that the simulation of $1/f^\beta$ noise with the desirable slope $\beta$ requires finding the special distributions of parameters of the system under consideration, at least a wide range of relaxation time constant should be assumed in order to correlate with the experiments (see also \cite{Reviews,Wong,Kaulakys2,Kaulakys2007}). 

Models of $1/f$ noise in some particular systems are usually specific and do not explain the omnipresence of processes with $1/f^\beta$ spectrum. Predominantly $1/f$ noise problem has been analyzed in conducting media, semiconductors, metals, electronic devices and other electronic systems \cite{Johnson,Reviews,Wong,Bernamont,Shtengel}. The topic of $1/f$ noise in such systems has been comprehensively reviewed \cite{Reviews}, even recently \cite{Wong}.
Nevertheless, despite numerous suggested models, the origin of flicker noise even there still remains an open issue: "More and more studies suggest that if there is a common regime for the low frequency noise, it must be mathematical rather than the physical one" \cite{Wong}. Here we can additionaly mention the disputed quantum theory \cite{Handel} of $1/f$ noise and satisfactorily interpreted $1/f$ noise in quantum chaos \cite{QChaos}.

In 1987 Bak, Tang, and Wiesenfeld \cite{Bak,Bak2} introduced the notion of self-organized criticality (SOC) with one of the main motivation to explain the universality of $1/f$ noise. 
SOC systems are nonequilibrium systems driven by their own dynamics to a self-organization. Fluctuations around this state, the so-called avalanches, are characterized by the power-law distributions in time and space, criticality implying long-range correlations. The distributions of avalanche sizes, durations and energies are all seen to be power laws. 

Two types of correlations should be distinguished in SOC: the scale-free distribution of their avalanche sizes and temporal correlations between avalanches, bursts or (rare, extreme) events. 
In the standard SOC models the search of $1/f^\beta$ noise is based on the observable power-law dependence of the burst size as a function of the burst duration and the power-law distribution of the burst sizes, with the Poisson distributed interevent times. Such power-laws usually result in the relatively high-frequency power-law, $1/f^\beta$, behavior of the power spectrum with the exponent $1.4\lesssim \beta \lesssim 2$ \cite{Jensen,Kuntz}. This mechanism of the power-law spectrum is related to the statistical models of $1/f^\beta$ noise representing signals as consisting of different random pulses \cite{Heiden,Schick,Ruseckas}. 

It should also be mentioned that originally SOC has been suggested as an explanation of the occurrence of 1/f noise and fractal pattern formation in the dynamical evolution of certain systems. However, recent research has revealed that the connection between these and SOC is rather loose \cite{Frigg}. 
Though an explanation of $1/f$ noise was one of the main motivations for the initial proposal of SOC, time dependent properties of self-organized critical systems had not been studied much theoretically so far \cite{Dhar}. 

It is of interest to note, that paper \cite{Bak} is the most cited paper in the field of $1/f$ noise problem, but it has been shown later on \cite{Jensen,Kuntz} that the proposed in Ref \cite{Bak} mechanism in SOC systems results in $1/f^\beta$ fluctuations with $1.5 \lesssim \beta \lesssim 2$ and does not explain the omnipresence of $1/f$ noise. On the other hand, we can point a recent paper \cite{Baiesi} where an example of $1/f$ noise in the classical sandpile model has been provided. 

It should be emphasized, however, that another mechanism of $1/f^\beta$ noise, based on the temporal correlations between avalanches, bursts or (rare, extreme) events, may be the source of the power-law $1/f^\beta$ spectra with $\beta \lesssim 1$ \cite{Davidsen}. Moreover, SOC is closely related with the observable $1/f^\beta$ crackling noise \cite{Crackling}, Barkhausen noise \cite{Barkhausen}, fluctuations of the long-term correlated seismic events \cite{Earthquakes} and $1/f^\beta$ fluctuations at non-equilibrium phase transitions \cite{Novak}. 

Ten years ago we proposed \cite{Kaulakys,Kaulakys1}, analyzed \cite{Kaulakys3} and later on generalized \cite{Kaulakys2} a simple point process model of $1/f^\beta$ noise and applied it for the financial systems \cite{Gontis}. Moreover, starting from the point process model we derived the stochastic nonlinear differential equations, i.e., the general Langevin equations with a multiplicative noise for the signal intensity exhibiting $1/f^\beta$ noise (with $0.5\le\beta\le 2$) in any desirable wide range of frequency $f$ \cite{Kaulakys4}.
Here we analyze the scaling properties of the signals generated by the particular stochastic differential equations. We obtain and analyze the power-law dependencies of the signal intensity, power spectrum, autocorrelation functions and the second order structural functions. The comparison with the numerical simulations is presented. 

Moreover, the numerical analysis reveals the second (reminder, that we start from the point process) structure of the signal composed of peaks, bursts, clusters of the events with the power-law distributed burst sizes $S$, burst durations $T$ and the inter-burst time $\theta$, while the burst sizes are approximately proportional to the squared durations of the bursts, $S \sim T^2$. Therefore, the proposed nonlinear stochastic model may simulate SOC and other similar systems where the processes consist of avalanches, bursts or clustering of the extreme events \cite{Bak,Bak2,Jensen,Kuntz,Frigg,Dhar,Baiesi,Davidsen,Crackling,Barkhausen,Earthquakes,Novak,Thurner}. 

\section{The model}
We start from the point process,
\begin{equation}
x(t)=a\sum_{k}\delta(t-t_k),
\label{deltasignal}
\end{equation}
representing the signal, current, flow, etc, $x(t)$, as a sequence of correlated pulses or series of events. Here $\delta(t)$ is the Dirac $\delta$-function and $a$ is a contribution to the signal $x(t)$ of one pulse at the time moment $t_k$. Our model is based on the generic multiplicative process for the interevent time $\tau_k=t_{k+1}-t_k$,
\begin{equation}
\tau_{k+1}=\tau_k+\gamma\tau_k^{2\mu-1}+\sigma\tau_k^\mu\varepsilon_k,
\label{recurrent}
\end{equation}
generating the power-law distributed,
\begin{equation}
P_k(\tau_k)\sim\tau_k^\alpha,\quad \alpha=\frac{2\gamma}{\sigma^2}-2\mu,
\label{dist}
\end{equation}
sequence of the interevent times $\tau_k$ \cite{Kaulakys2,Gontis}. 

Some motivations for equation \eref{recurrent} were given on papers \cite{Kaulakys,Kaulakys2,Kaulakys3,Gontis}. Additional comments are presented below, after equation \eref{Poisson}. 

Therefore, in our model the (average) interevent time $\tau_k$ fluctuates due to the random perturbations by a sequence of uncorrelated normally distributed random variables $\{\varepsilon_k\}$ with zero expectation and unit variance, $\sigma$ is the standard deviation of the white noise and $\gamma\ll 1$ is a coefficient of the nonlinear damping.

Transition from the occurrence number $k$ to the actual time $t$ in equation \eref{recurrent} according to the relation $dt=\tau_kdk$ yields the It\^o stochastic differential equation for the variable $\tau(t)$ as a function of the actual time,
\begin{equation}
\mathrm{d}\tau=\gamma\tau^{2\mu-2}\mathrm{d}t+\sigma\tau^{\mu-1/2}\mathrm{d}W,
\label{time}
\end{equation}
where $W$ is a standard Wiener process. Equation \eref{time} generates the stochastic variable $\tau$, power-law distributed,
\begin{equation}
P_t(\tau)=\frac{P_k(\tau)}{\left<\tau_k\right>} \tau \sim\tau^{\alpha+1},
\label{taudist}
\end{equation}
in the actual time $t$. Here $\left<\tau_k\right>$ is the average interevent time. $\tau(t)$ may be interpreted as the average time-dependent interevent time of the modulated Poisson-like process with the distribution of the interevent time 
\begin{equation}
P_p(\tau_p)=\frac{1}{\tau\left(t\right)}\mathrm{e}^{-\tau _p/\tau \left( t\right) }=n\left( t\right) \mathrm{e}^{-n\left( t\right) \tau _p}, 
\label{Poisson}
\end{equation}
where $n(t)=1/\tau(t)$ is the time dependent rate of the process \cite{Gontis}. 

Additional support for the stochastic model \eref{deltasignal} -- \eref{Poisson} of the scaled processes and $1/f^\beta$ noise is the following. The fluctuations of the intensity of the signals, currents, flows, etc, consisting of the discrete objects (electrons, photons, packets, vehicles, pulses, events, etc) are primarily and basically defined by fluctuations of the (average) interevent, interpulse, interarrival, recurrence, or waiting time. Equation \eref{time} is a special case of the general non-linear Langevin equation 
\begin{equation}\mathrm{d}\tau = d(\tau)\mathrm{d}t + b(\tau)\mathrm{d}W(t)\label{Langevin}
\end{equation}
with the drift coefficient $d(\tau)$ and a multiplicative noise $b(\tau)\xi(t)$ for the (average) interevent time $\tau(t)$, with $\xi(t)$ being a white nose defined from the relation $\mathrm{d}W(t) = \xi(t)\mathrm{d}(t)$. 
Equation \eref{Langevin} is a straight analogy of the well-known Langevin equation for the continuous random variable $x$. For the process consisting of the discrete objects the intensity of the signal fluctuates due to fluctuations of the rate, i.e., density of the objects in the time axis, which is a consequence of fluctuations of the interarrival or interevent time. Equation \eref{Langevin} in reality represents (in the simplest form) such fluctuations due to random perturbations by white noise. 

In papers \cite{Kaulakys,Kaulakys1,Kaulakys2} it has been shown that the small interevent times and clustering of the events make the greatest contribution to $1/f^\beta$ noise, low frequency fluctuations and exhibition of the long-range scaled features. Therefore, it is straight to approximate the non-linear diffusion coefficient $b(\tau)$ and the distribution of the interevent time $P_t(\tau)$ in some interval of small interevent times $\tau$ by the power-law dependences or expensions, \begin{equation}b(\tau) = \sigma \tau^{\mu-\frac{1}{2}}, 	
\label{b}
\end{equation}\begin{equation}P_t(\tau)\sim\tau^{\alpha+1}.\label{tau_dist2}
\end{equation}

The power-law distribution of the interevent, recurrence, or waiting time is observable in different systems from physics and seismology to the Internet, financial markets and neural spikes (see, e.g., \cite{Kaulakys2,Gontis,Thurner}).
It should be noted that the multiplicative equations with the drift coefficient $d(\tau)$ proportional to the Stratonovich correction for the drift, leading the transformation from the Stratonovich to the It\^o stochastic differential equation \cite{Arnold}, i.e., when 
\begin{equation}
d(\tau) \sim  \frac{1}{2}b(\tau)b^\prime(\tau), \label{d}
\end{equation}
with the power-law depending, like \eref{b}, diffusion coefficient $b(\tau)$, generates the power-law distribution of the stochastic variable. Equations \eref{recurrent} and \eref{time} are definitely of such kind. 
Therefore, equation \eref{time} is one of the simplest multiplicative equations for the interevent time, modeling scaled processes, while equation (2) is just the lowest order difference equation following from equation \eref{time} when the step of integration $\Delta t_k$ equals the interevent time $\tau_k$. 

The It\^o transformation in equation \eref{time} of the variable from $\tau$ to the averaged over the time interval $\tau$ intensity of the signal $x(t)=a/\tau(t)$ \cite{Kaulakys4} yields the class of It\^o stochastic differential equations
\begin{equation}
\mathrm{d}x=\left(\eta-\frac{1}{2}\lambda\right)x^{2\eta-1}\mathrm{d}t_s+x^\eta \mathrm{d}W
\label{ito}
\end{equation}
for the signal as function of the scaled time
\begin{equation}
t_s=\frac{\sigma^2}{a^{3-2\mu}}t.
\label{scaledtime}
\end{equation}
Here the new parameters
\begin{equation}
\eta=\frac{5}{2}-\mu,\quad\lambda=3+\alpha=\frac{2\gamma}{\sigma^2}+2(\eta-1) 
\label{param}
\end{equation}
have been introduced.

The Fokker-Plank equation associated with equation \eref{ito} gives the power-law distribution density of the signal intensity
\begin{equation}
P(x)\sim\frac{1}{x^\lambda}
\label{sigdistr}
\end{equation}
with the exponent $\lambda$.

For $\lambda>1$ distribution \eref{sigdistr} diverge as $x\to 0$, and, therefore, the diffusion of $x$ should be restricted at least from the side of small values, or equation \eref{ito} should be modified. Thus, further we will consider the modified equation for $x>0$ only, 
\begin{equation}
\mathrm{d}x=\left(\eta-\frac{1}{2}\lambda\right)(x_m+x)^{2\eta-1}\mathrm{d}t_s+(x_m+x)^\eta \mathrm{d}W,
\label{signal}
\end{equation}
with the additional small parameter $x_m$ restricting the divergence of the power-law distribution of $x$ at $x=0$.

Equation \eref{signal} for small $x\ll x_m$ represents the linear additive stochastic process generating the Brownian motion with the steady drift, while for $x\gg x_m$ it reduces to the multiplicative equation \eref{ito}.

\section{Analysis of the model}
The Fokker-Plank equation associated with equation \eref{signal} gives the steady-state solution for distribution of $x$,
\begin{equation}
P(x)=\frac{(\lambda-1)x_m^{\lambda-1}}{(x_m+x)^\lambda},\quad x>0,\quad\lambda>1.
\label{steadydist}
\end{equation}
We can obtain the power spectral density of the signal generated by equation \eref{signal} from equation (28) derived in paper \cite{Kaulakys2}. After some algebra we can write
\begin{equation}
S(f)=\frac{A}{f^\beta},\quad f\gg f_1=\frac{2+\lambda-2\eta}{2\pi}x_m^{2(\eta-1)}
\label{spectrum}
\end{equation}
with
\begin{equation}
A=\frac{(\lambda-1)\Gamma(\beta-1/2)x_m^{\lambda-1}}{2\sqrt{\pi}(\eta-1)\sin(\pi\beta/2)}\left(\frac{2+\lambda-2\eta}{2\pi}\right)^{\beta-1}
\label{consta}
\end{equation}
and
\begin{equation}
\beta=1+\frac{\lambda-3}{2(\eta-1)}
\label{beta}
\end{equation}
for $0.5<\beta<2$, $4-\eta<\lambda<1+2\eta$ and $\eta>1$. Note that the frequency $f$ in equation \eref{spectrum} is the scaled frequency matching the scaled time $t_s$ \eref{scaledtime}.

The autocorrelation function $C(s)$ of the process can be expressed according to Wiener-Khinchin theorem as the inverse Fourier transform of the power spectrum,
\begin{equation}
C(s)=\left<x(t)x(t+s)\right>=\int_0^\infty S(f)\cos(2\pi fs)\mathrm{d}f.
\label{correlation}
\end{equation}

A pure $1/f^\beta$ power spectrum is physically impossible because of the total power would be infinity. Depending on whether $\beta$ is greater or less than one it is necessary to introduce a low frequency cutoff $f_{\min}$ or a high frequency cutoff $f_{\max}$ \cite{Theiler,Talocia}. For calculation of the autocorrelation function according to equation \eref{correlation} when $\beta>0$ it is not necessary to introduce the hight frequency cutoff. 

Usually one introduces a discontinuous transition to the flat spectrum at the lower cutoff $f_{\min}$ \cite{Theiler,Talocia}. 
Here at low frequencies we will insert the smooth transition to the flat spectrum in the vicinity of $f_0 \sim f_{\min}$, i.e., we will approximate the power spectrum \eref{spectrum} as 
\begin{equation}
S(f)=\frac{A}{(f_0^2+f^2)^{\beta/2}}=
\cases{A/f_0^\beta, \qquad f\ll f_0, \\ A/f^\beta, \qquad f\gg f_0.}
\label{spectrum_aprox} 
\end{equation}
Inserting \eref{spectrum_aprox} into equation \eref{correlation} we obtain 
\begin{equation}
C(s)=\frac{A\sqrt{\pi}}{\Gamma(\beta/2)f_0^{\beta-1}}\left(\frac{z}{2}\right)^hK_h(z),
\label{correlation_aprox}
\end{equation}
where $K_h(z)$ is the modified Bessel function, $z=2\pi f_0s$ and $h=(\beta-1)/2$.

The first two terms of expansion of equation \eref{correlation_aprox} in powers of $z$ are
\begin{equation}
C(s)=\frac{A\sqrt{\pi}}{2\Gamma(\beta/2)f_0^{\beta-1}}\left[\Gamma\left(\frac{\beta-1}{2}\right)+\Gamma\left(\frac{1-\beta}{2}\right)(\pi f_0s)^{\beta-1}\right] 
\label{correlation_expansion}
\end{equation}
for $h\neq 0$, i.e, $\beta\neq 1$, and 
\begin{equation}
C(s)=AK_0(2\pi f_0s)\simeq A[-\gamma-\ln(\pi f_0s)]=\mathrm{const}-A\ln s
\label{correlation_expansion_0}
\end{equation}
for $h=0$, i.e., for the pure $1/f$ noise with $\beta=1$. Here $\gamma\simeq 0.577216$ is Euler's constant.

The leading terms of expression \eref{correlation_expansion} are different, depending on whether $\beta<1$ or $\beta>1$. Thus for $h<0$, i.e., when $0<\beta<1$
\begin{equation}
C(s)=\frac{A\Gamma(1-\beta)}{(2\pi s)^{1-\beta}}\sin\left(\frac{\pi\beta}{2}\right)\sim\frac{1}{s^{1-\beta}}, 
\label{correlation_h_less}
\end{equation}
while for $h>0$, i.e., for $1<\beta<3$
\begin{equation}
C(s)=C(0)-Bs^{\beta-1}.
\label{correlation_h_greater}
\end{equation}
Here
\begin{equation}
C(0)=\left<x^2\right>=\int_0^\infty S(f)\mathrm{d}f=\frac{A\sqrt{\pi}\Gamma\left(\frac{\beta-1}{2}\right)}{2f_0^{\beta-1}\Gamma\left(\frac{\beta}{2}\right)}
\label{correlation_0}
\end{equation}
and 
\begin{equation}
B=\frac{A\pi^{\beta+\frac12}}{2\Gamma\left(\frac{\beta}{2}\right)\Gamma\left(\frac{\beta+1}{2}\right)\sin\left(\pi\frac{\beta-1}{2}\right)}=-A(2\pi)^{\beta-1}\Gamma(1-\beta)\sin\left(\frac{\pi\beta}{2}\right).
\label{B}
\end{equation}
For $\beta=2$ equations \eref{correlation_aprox} and \eref{correlation_h_greater} -- \eref{B} yield
\begin{equation}
C(s)=C(0)e^{-2\pi f_0s}=\frac{A\pi}{2f_0}e^{-2\pi f_0s}=C(0)-A\pi^2s\pm...
\label{correlation_beta_2}
\end{equation}

It should be noted, that particular cases \eref{correlation_expansion_0} -- \eref{correlation_beta_2} of the general expressions \eref{correlation_aprox} and \eref{correlation_expansion} are in agreement with the results of papers \cite{Theiler,Talocia,Caprari} obtained with the non-uniform cutoff of the spectrum at low frequency. On the other hand, the introduced parameter $h$ for $\beta\geq1$ coincides with the Hurst exponent $H$ \cite{Talocia},
\begin{equation}
H=\cases{0, \qquad 0<\beta\leq1,\\ \frac{1}{2}(\beta-1), \qquad 1<\beta<3, \\ 1, \qquad 3\leq\beta<4.}
\label{hurst}
\end{equation}

The exponent $H$ is associated with the scaling of the second order structural function, or height-height correlation function \cite{Theiler,Talocia,Caprari,Kaulakys5}
\begin{equation}
F(s)=F_2^2(s)=\left<\left|x(t+s)-x(t)\right|^2\right>\sim s^{2H}.
\label{height-height}
\end{equation}
The exponent $H$ characterizes the power-law diffusion rate, as well. This variance of the differenced time series (delayed signal) may be expressed as 
\begin{equation}
F(s)=\left<x^2(t+s)\right>+\left<x^2(t)\right>-2\left<x(t+s)x(t)\right>=2[C(0)-C(s)].
\label{height-height2}
\end{equation}
Substituting expressions \eref{correlation} and \eref{correlation_0} into \eref{height-height2} we have
\begin{equation}
F(s)=F_2^2(s)=4\int_0^\infty S(f)\sin^2(\pi sf)\mathrm{d}f.
\label{height-height3}
\end{equation}
For the convergence of integral in \eref{height-height3} at $\beta\leq 1$ we need to cut off the power-law spectrum \eref{spectrum} at high frequency $f_{\max}$. Then the leading terms of the height-height correlation function \eref{height-height3} are
\begin{equation}
F(s)=2A\times\cases{\frac{f_{\max}^{1-\beta}}{1-\beta}\left[1-\frac{\Gamma(2-\beta)}{(2\pi f_{\max}s)^{1-\beta}}\sin\left(\frac{\pi\beta}{2}\right)\right], \qquad 0<\beta<1,\\ [\ln(\pi f_{\max}s)-\gamma], \qquad \beta=1.} 
\label{height-height4}
\end{equation}
For $1<\beta<3$ the integral in \eref{height-height3} may be integrated exactly and we have
\begin{equation}
F(s)=-2A\Gamma(1-\beta)\sin\left(\frac{\pi\beta}{2}\right)(2\pi s)^{\beta-1}=2Bs^{\beta-1},\quad 1<\beta<3.
\label{height-height5}
\end{equation}

The spatial power spectrum and the height-height correlation function \eref{height-height} are used for analysis of rough self-affine surfaces and assessing the growth mechanism of thin films \cite{Barabasi,Yang,Zhao,Yanguas-Gil,Palasantzas}, as well. There sometimes the violation of the scaling relation $\beta=2H+1$ is observable \cite{Yang,Zhao,Timashev}.

\begin{figure}[htb]
\centering
\includegraphics[width=.4\textwidth]{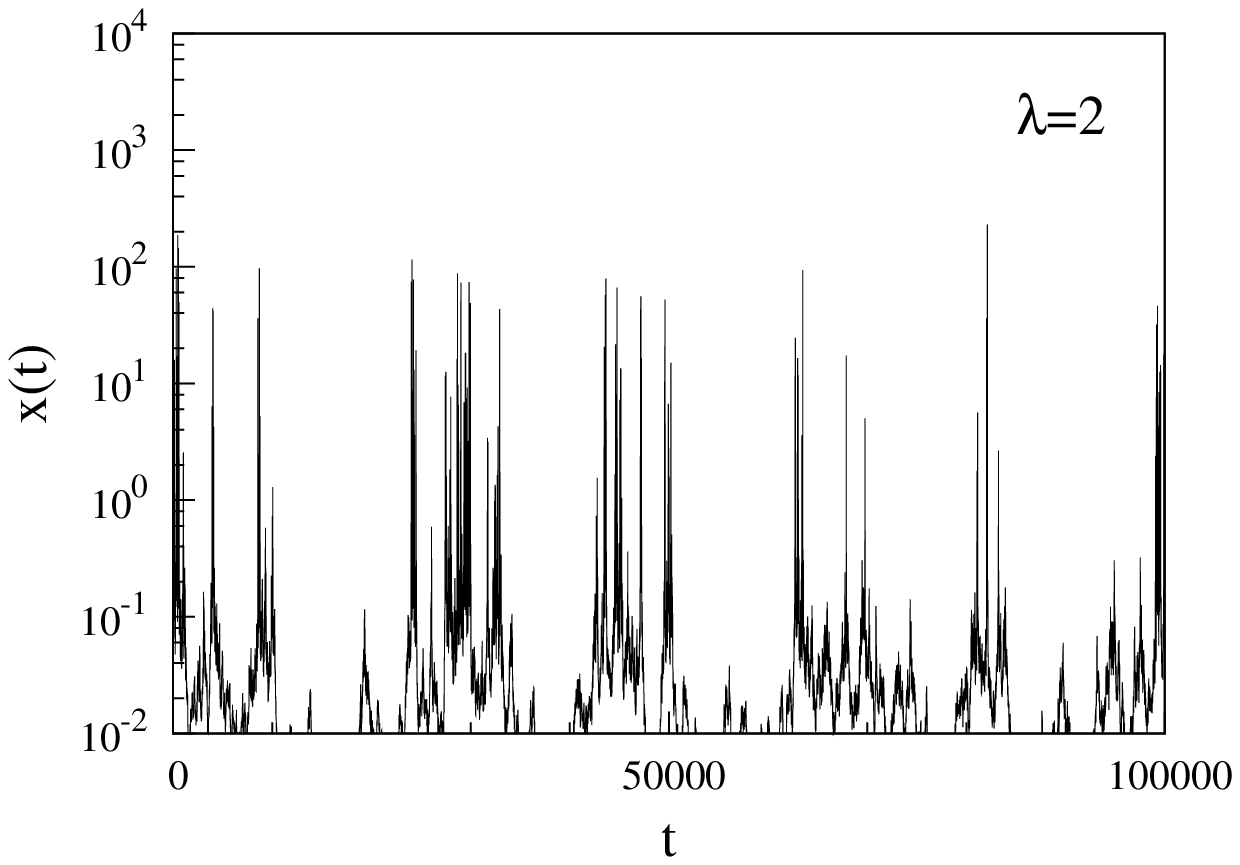}
\includegraphics[width=.4\textwidth]{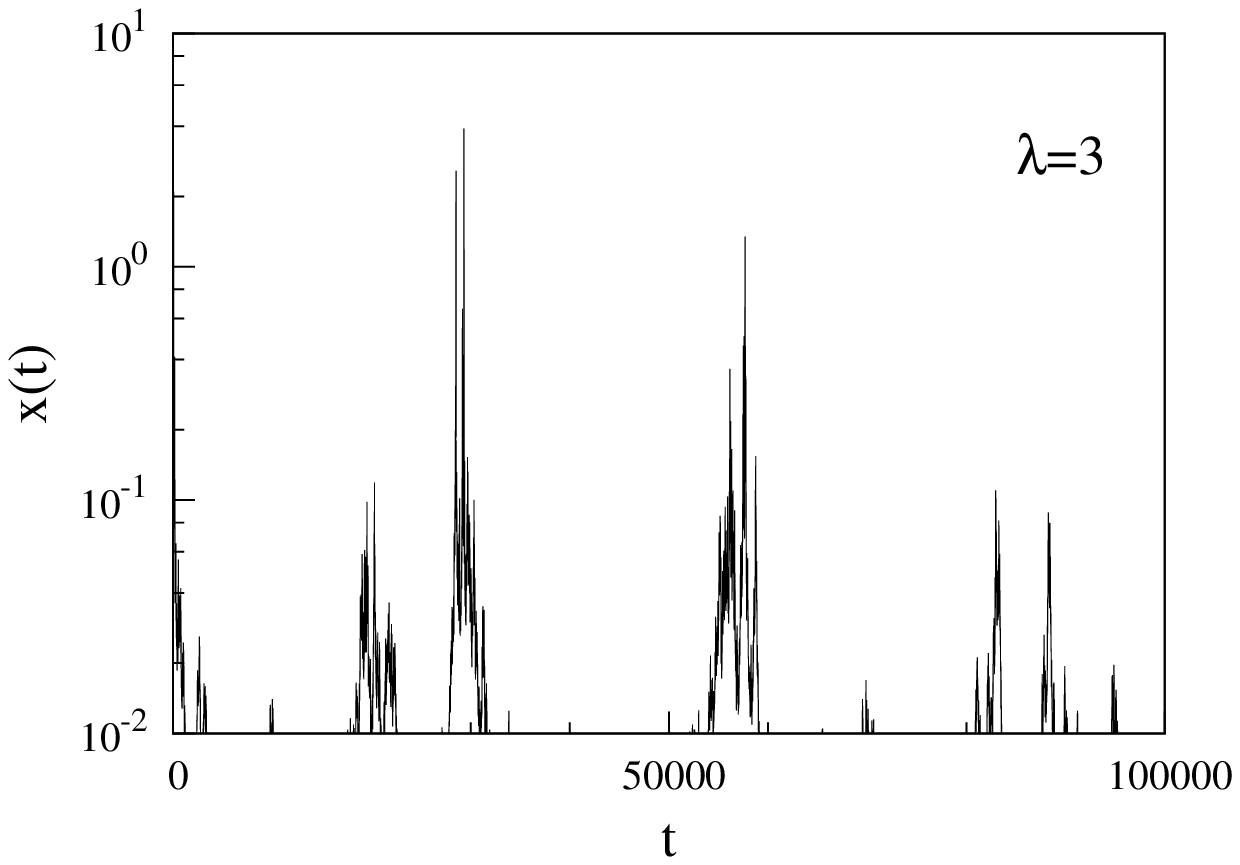}
\includegraphics[width=.4\textwidth]{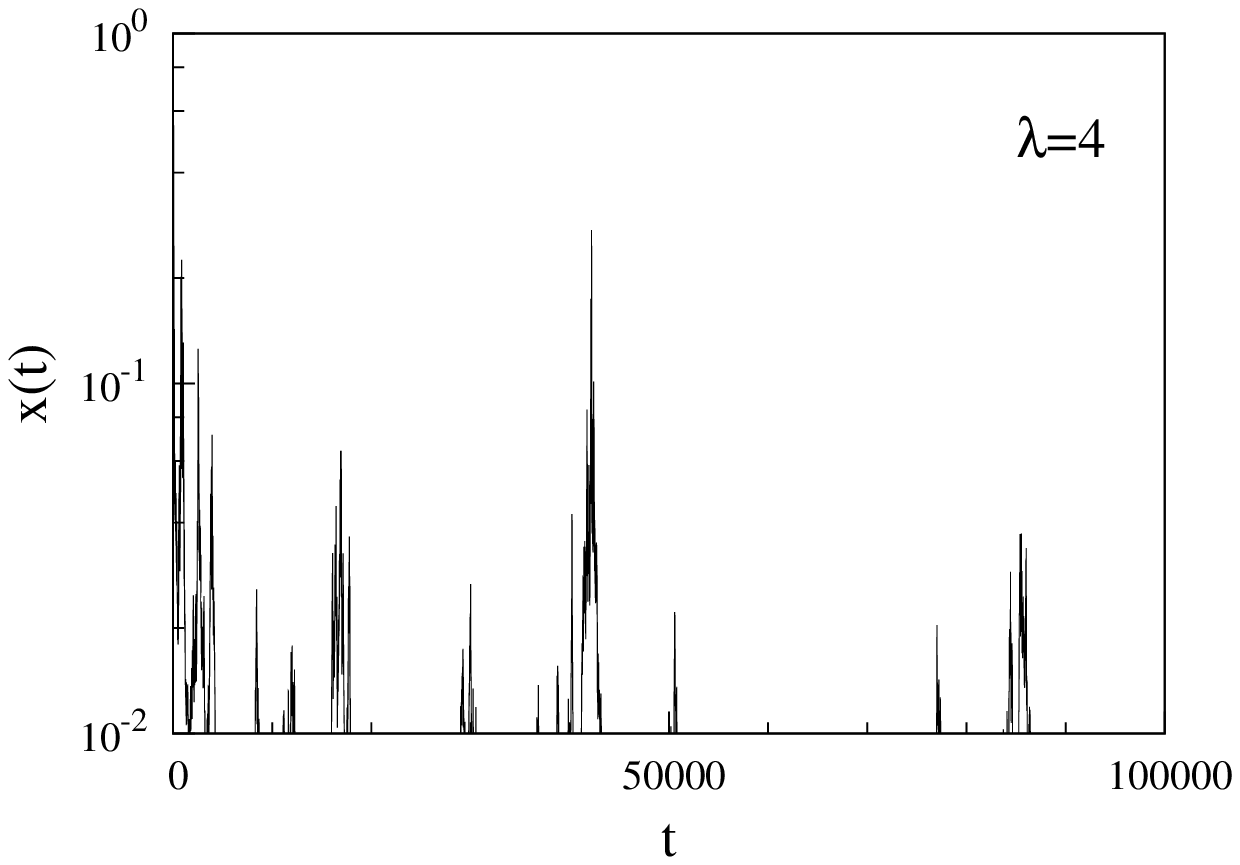}
\includegraphics[width=.4\textwidth]{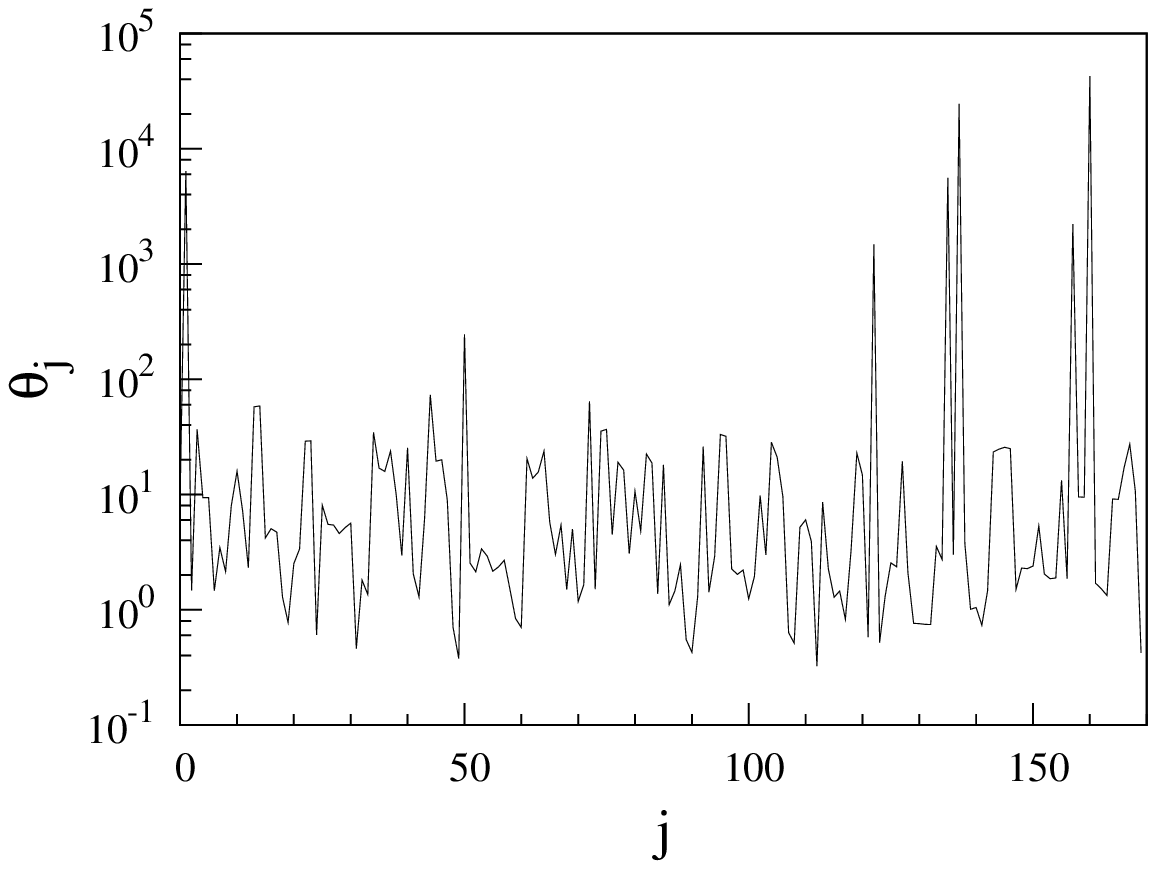}
\caption{Examples of the numerically computed signals according to equations \eref{discrete} and \eref{time2} with the parameters $\eta =2$, $x_m=10^{-2}$ and different values of $\beta$ and $\lambda$: $\beta=1/2$ when $\lambda=2$, $\beta=1$ when $\lambda=3$ and $\beta=3/2$ when $\lambda=4$ and the interburst time $\theta_j$ as a function of the occurrence number $j$ of the events peaking above the threshold value $x_{th} = 0.1$ for the pure $1/f$ noise with $\beta=1$ when $\lambda=3$.}
\label{figure_1}
\end{figure}

\begin{figure}[htb]
\centering
\includegraphics[width=.4\textwidth]{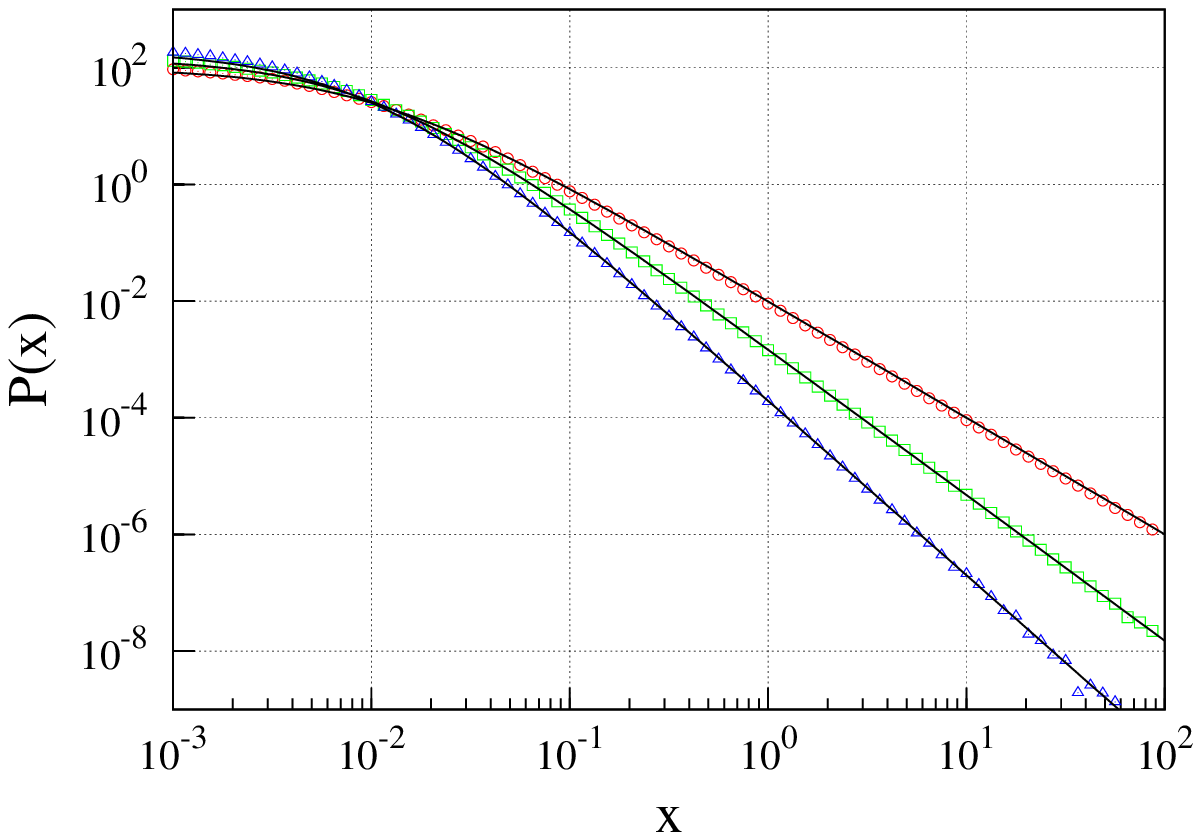}
\includegraphics[width=.4\textwidth]{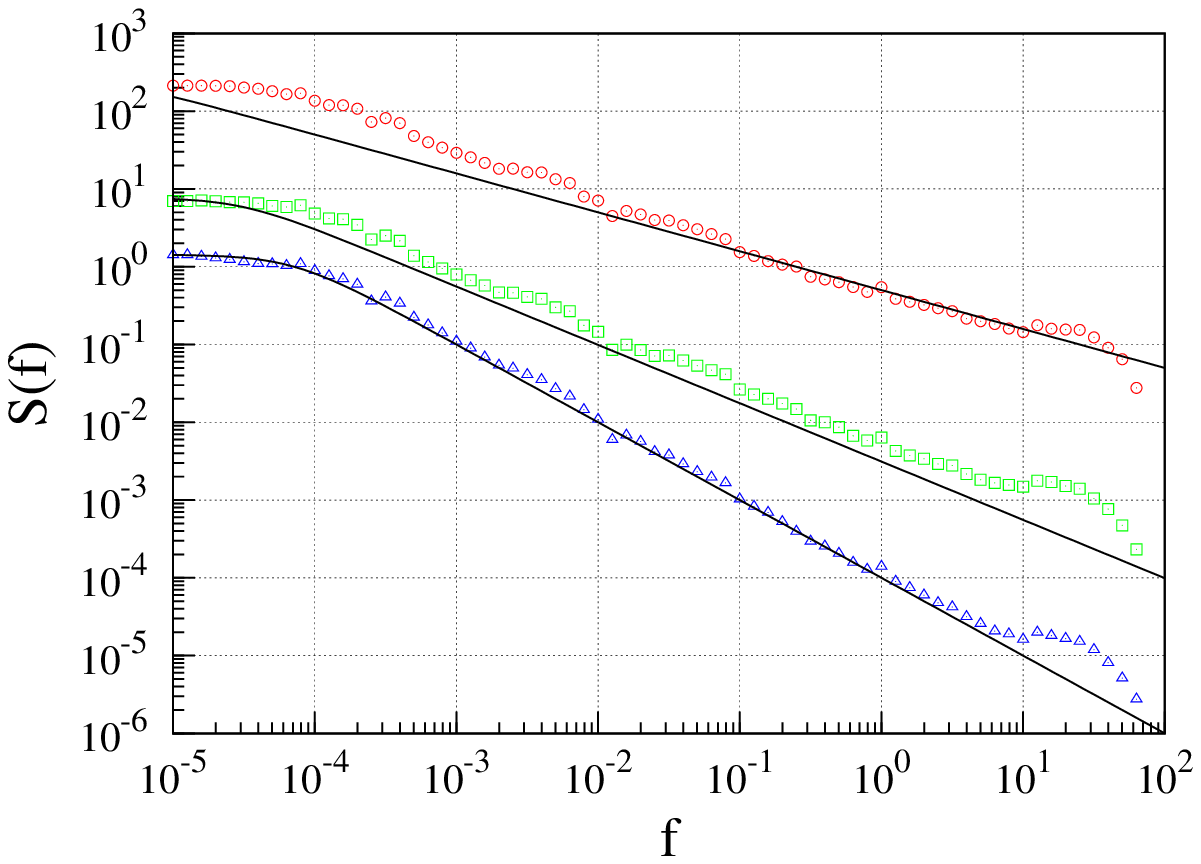}
\includegraphics[width=.4\textwidth]{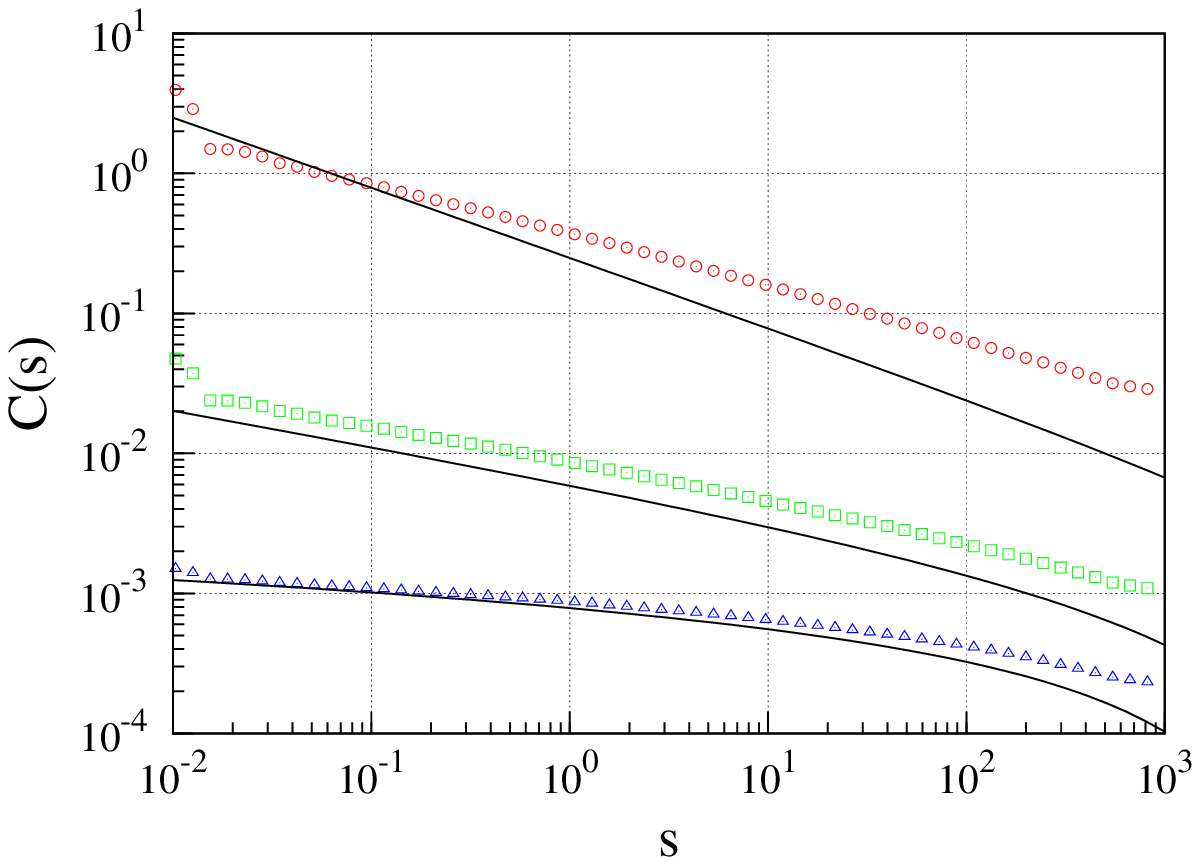}
\includegraphics[width=.4\textwidth]{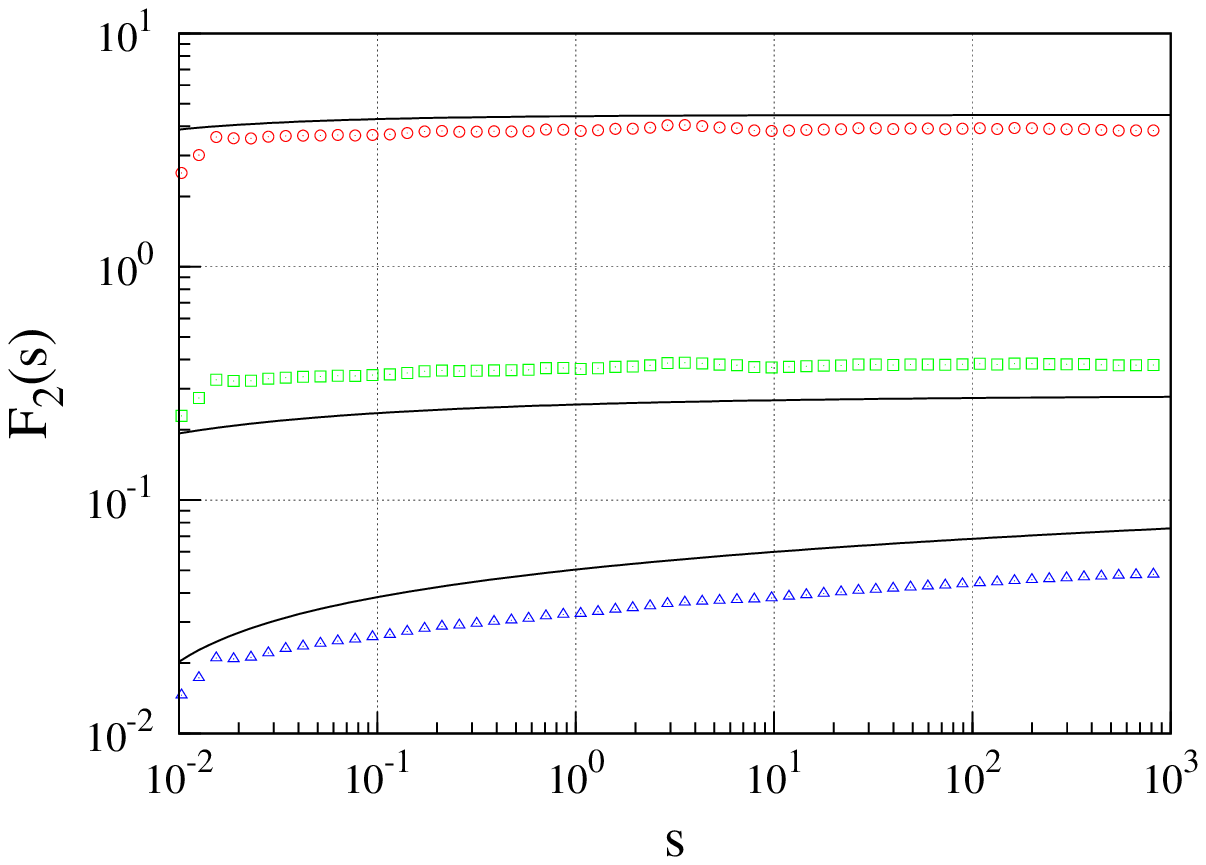}
\caption{Distribution density, $P\left( x\right)$, power spectral density, $S\left(f\right)$, autocorrelation function, $C(s)$, and the second order structural function, $F_2\left( s\right)$, for solutions of equation \eref{signal} with $\eta =2$, $x_m=10^{-2}$ and different values of the parameter $\lambda$: $\lambda=2$ (circles), $\lambda=2.5$ (squares) and $\lambda=3$ (triangles) in comparison with the analytical results (solid lines) according to equations \eref{steadydist}, \eref{consta} -- \eref{correlation_aprox}, \eref{height-height4} and \eref{height-height5}, respectively.}
\label{figure_2}
\end{figure}

\begin{figure}[htb]
\centering
\includegraphics[width=.4\textwidth]{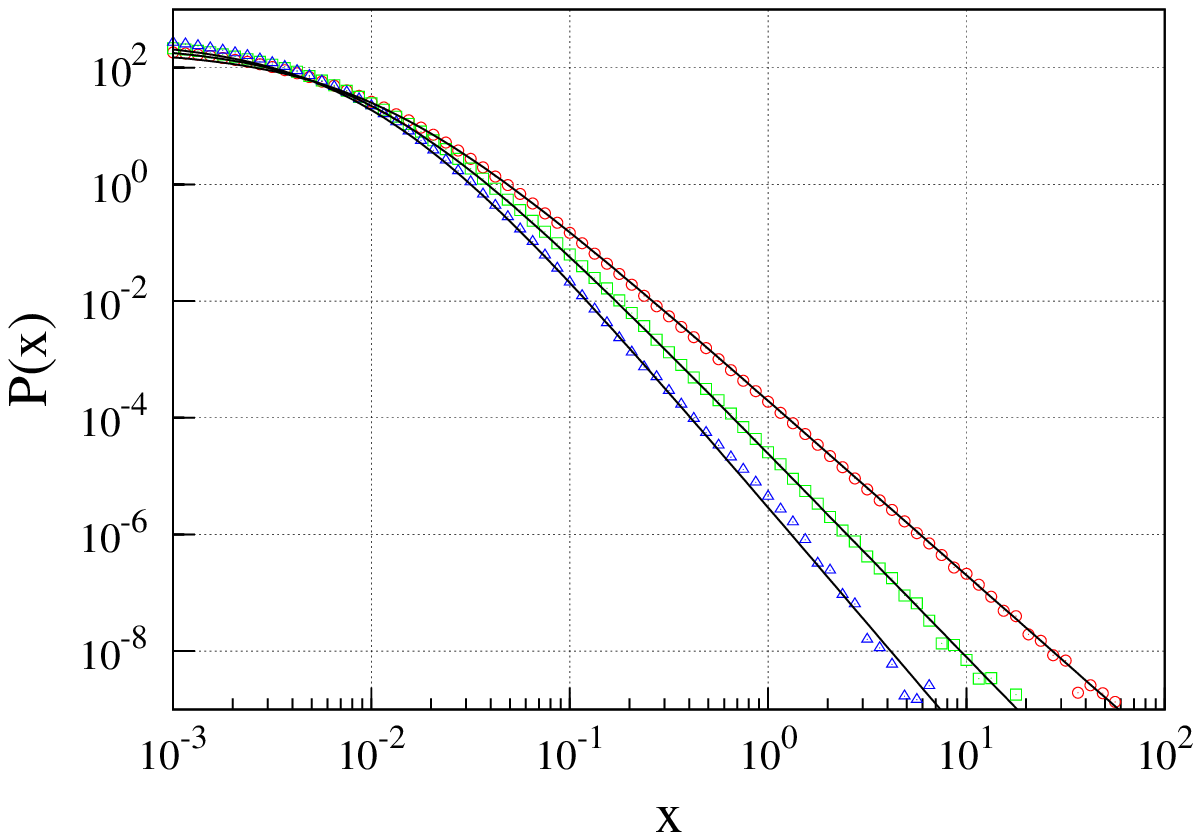}
\includegraphics[width=.4\textwidth]{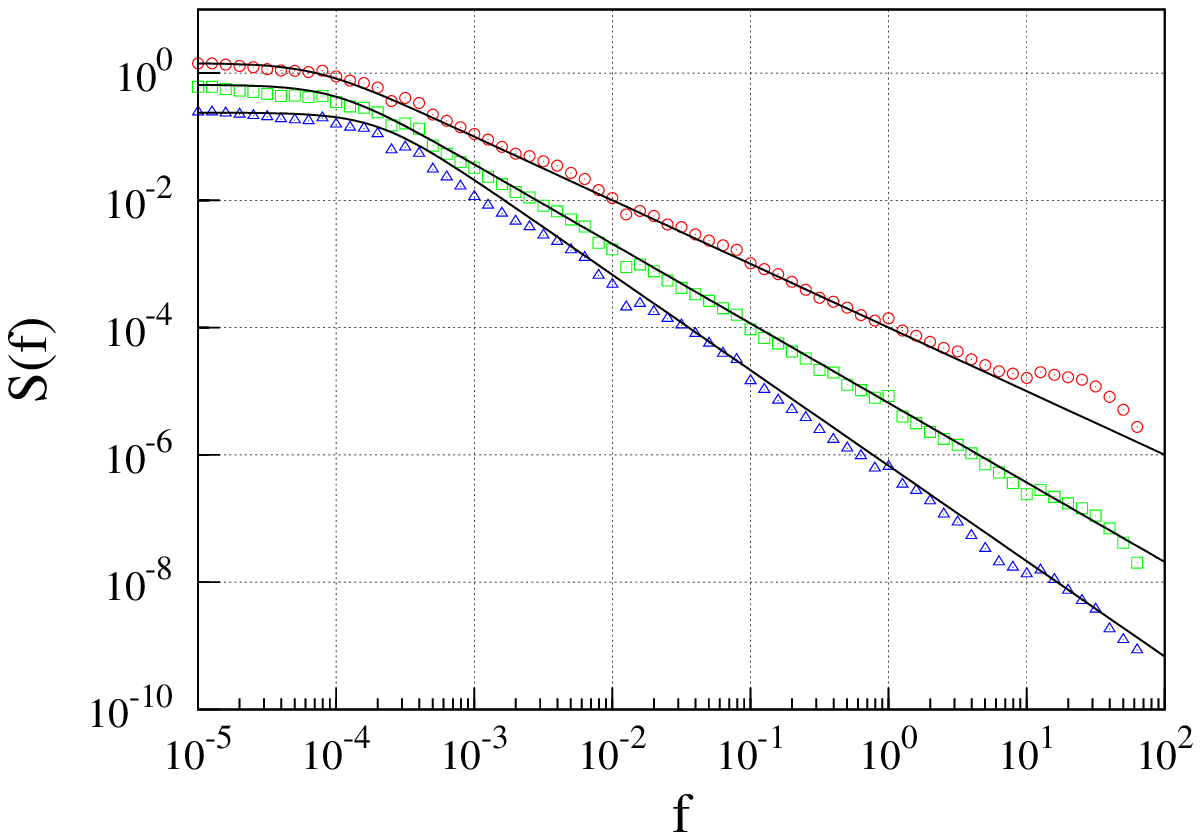}
\includegraphics[width=.4\textwidth]{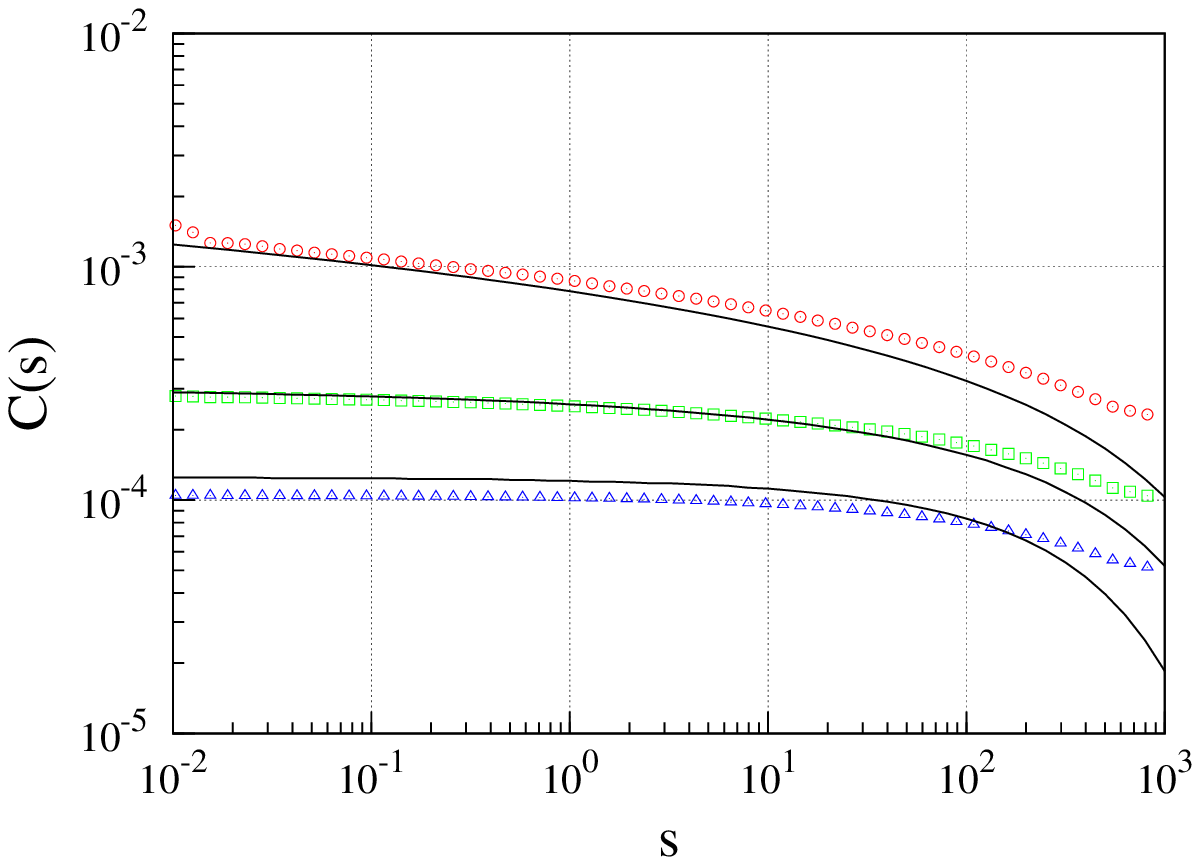}
\includegraphics[width=.4\textwidth]{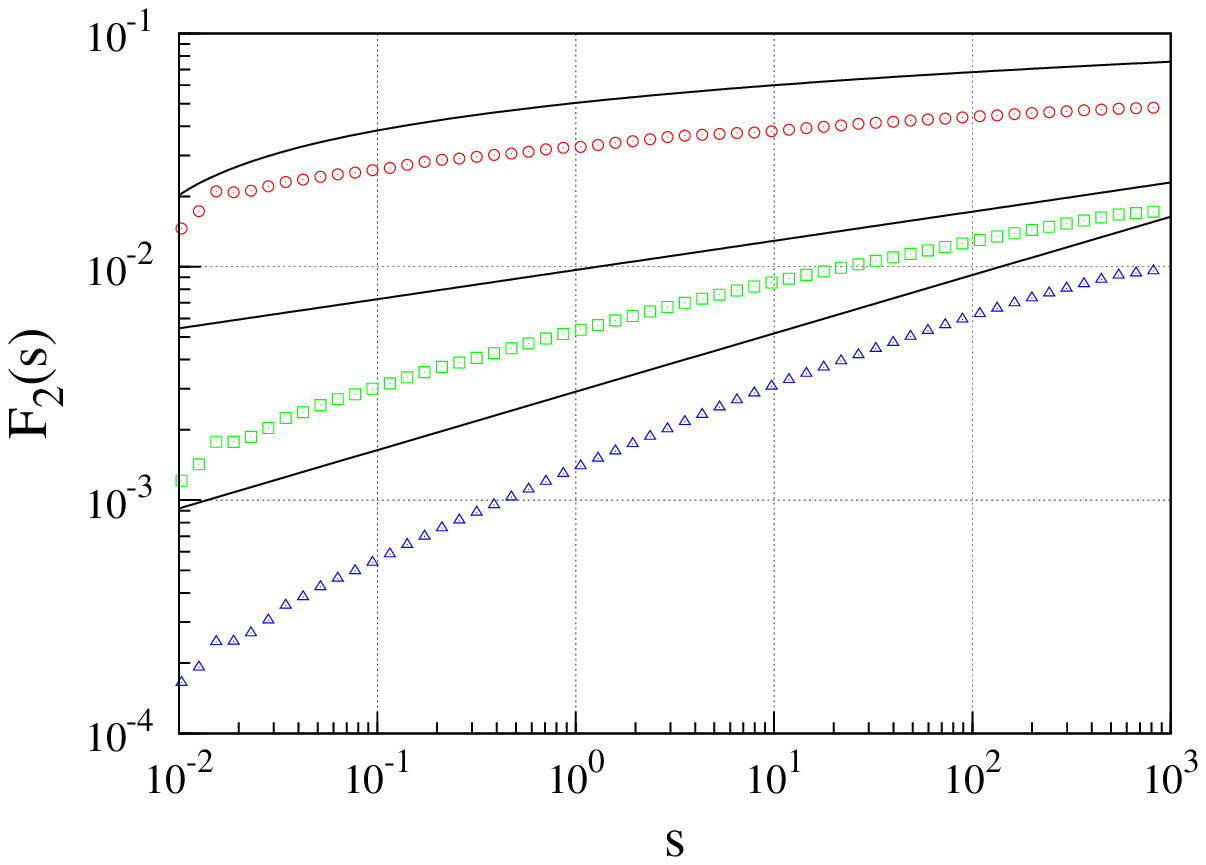}
\caption{As in figure \ref{figure_2} but for the parameters $\lambda=3$ (circles), $\lambda=3.5$ (squares) and $\lambda=4$ (triangles).}
\label{figure_3}
\end{figure}

\section{Numerical analysis}
For the numerical analysis we have to solve equation \eref{signal} and analyze the obtained numerical solutions. 
We can solve equation \eref{signal} using the method of discretization with the variable step of integration 
\begin{equation}
h_i=\Delta t_i=\frac{\kappa ^2}{\left( x_m+x_i\right) ^l}, \label{step}
\end{equation}
where $\kappa $ is a small parameter while the exponent $l$ rules the dependence of the integration step on the value of the variable $x$. Thus, $ l=0$ corresponds to the fixed step, for $l=1$ we have analogy with equation \eref{recurrent} when the step is proportional to the interevent time $\tau _k$, $l=2\left( \eta -1\right) $ matches the case when the change of the variable $x$ in one step is proportional to the value of the variable at time of the
step \cite{Kaulakys4} and so on. As a result we have the system of the difference equations 
\begin{equation}
x_{i+1}=x_i+\kappa ^2\left( \eta -\frac 12\lambda \right) \left(x_m+x_i\right) ^{2\eta -1-l}+\kappa \left(x_m+x_i\right) ^{\eta-l/2}\varepsilon _i, \label{discrete}
\end{equation}
\begin{equation}
t_{i+1}=t_i+\frac{\kappa ^2}{\left( x_m+x_i\right) ^l}, \quad x_i>0. \label{time2}
\end{equation}
Here $\varepsilon _i$ is a set of uncorrelated normally distributed random variables with zero expectation and unit variance. In the Milstein approximation equation \eref{discrete} should be replaced by the equation 
\begin{equation}
\fl x_{i+1}=x_i+\frac{\kappa ^2}2\left( \eta -\lambda \right) \left(x_m+x_i\right) ^{2\eta -1-l} +\kappa \left(_m+x_i\right) ^{\eta-l/2}\varepsilon _i +\frac{\kappa ^2\eta }2\left( x_m+x_i\right) ^{2\eta-1-l} 
\varepsilon _i^2. \label{Milstein}
\end{equation}

\begin{figure}[htb]
\centering
\includegraphics[width=.4\textwidth]{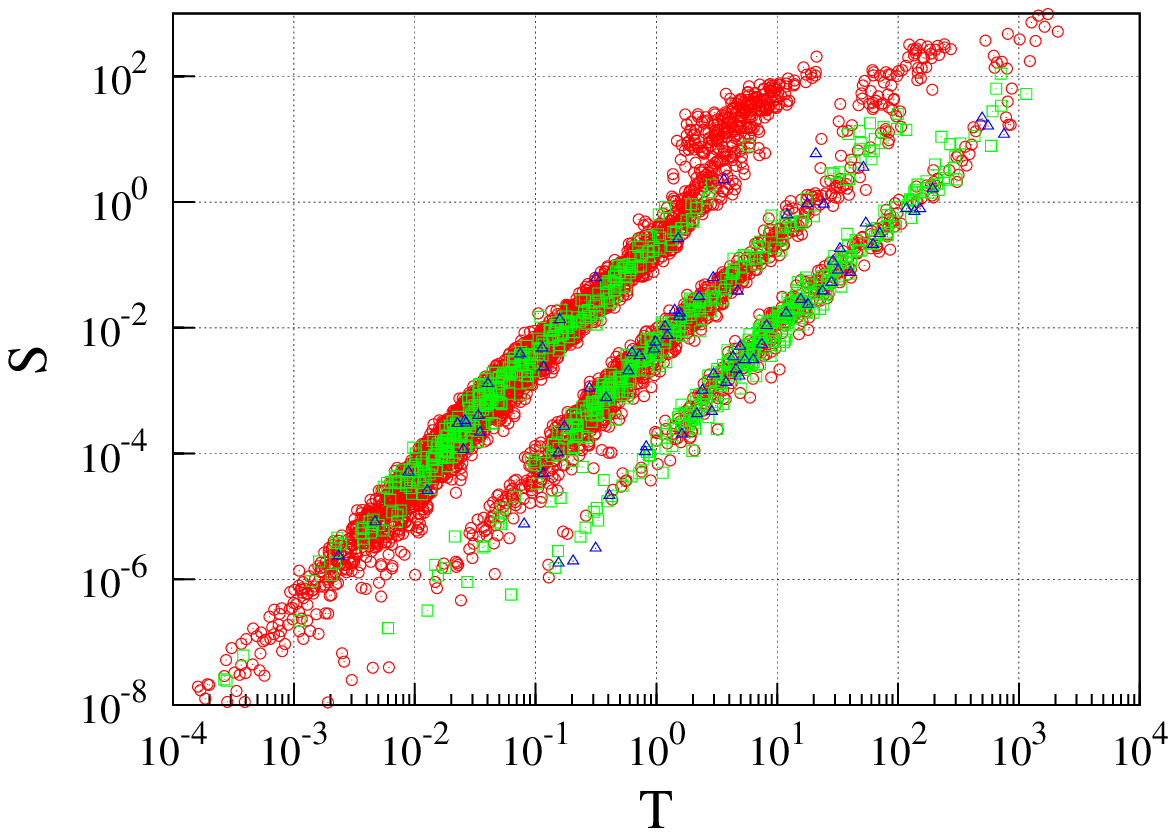}
\includegraphics[width=.4\textwidth]{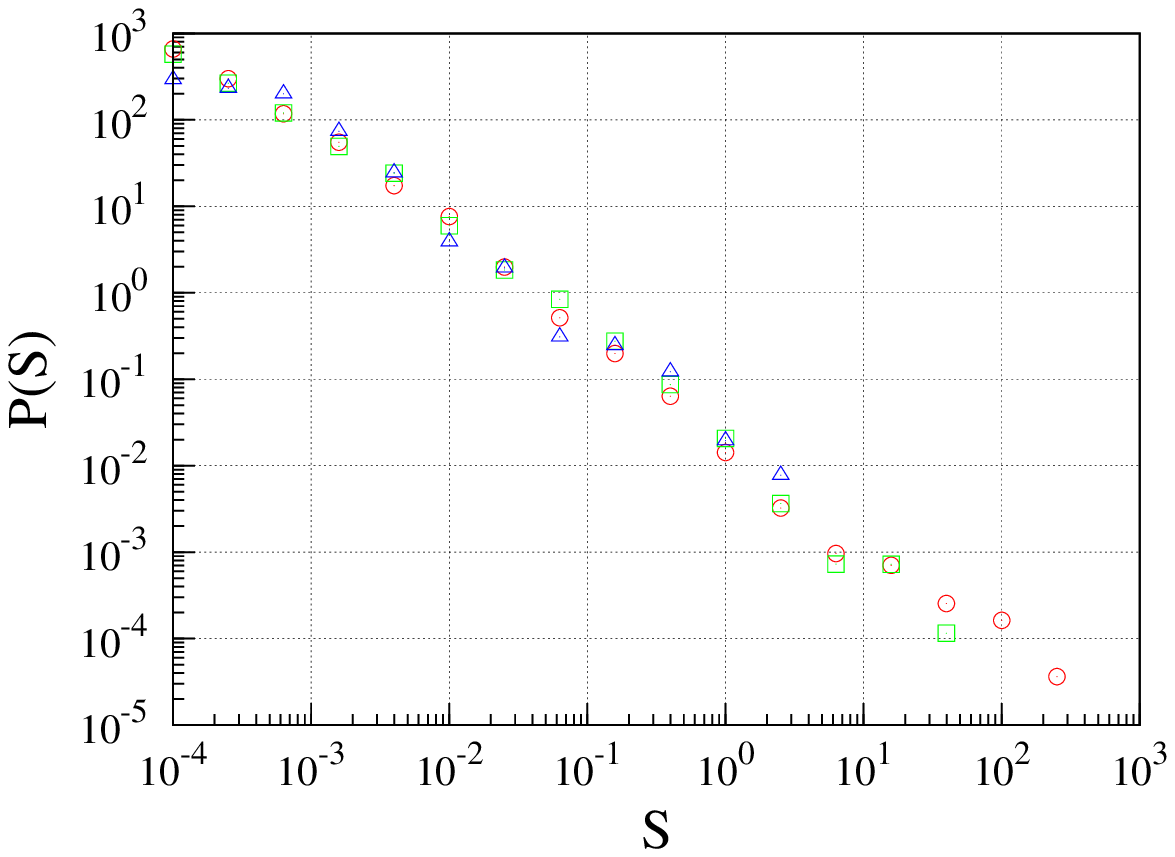}
\includegraphics[width=.4\textwidth]{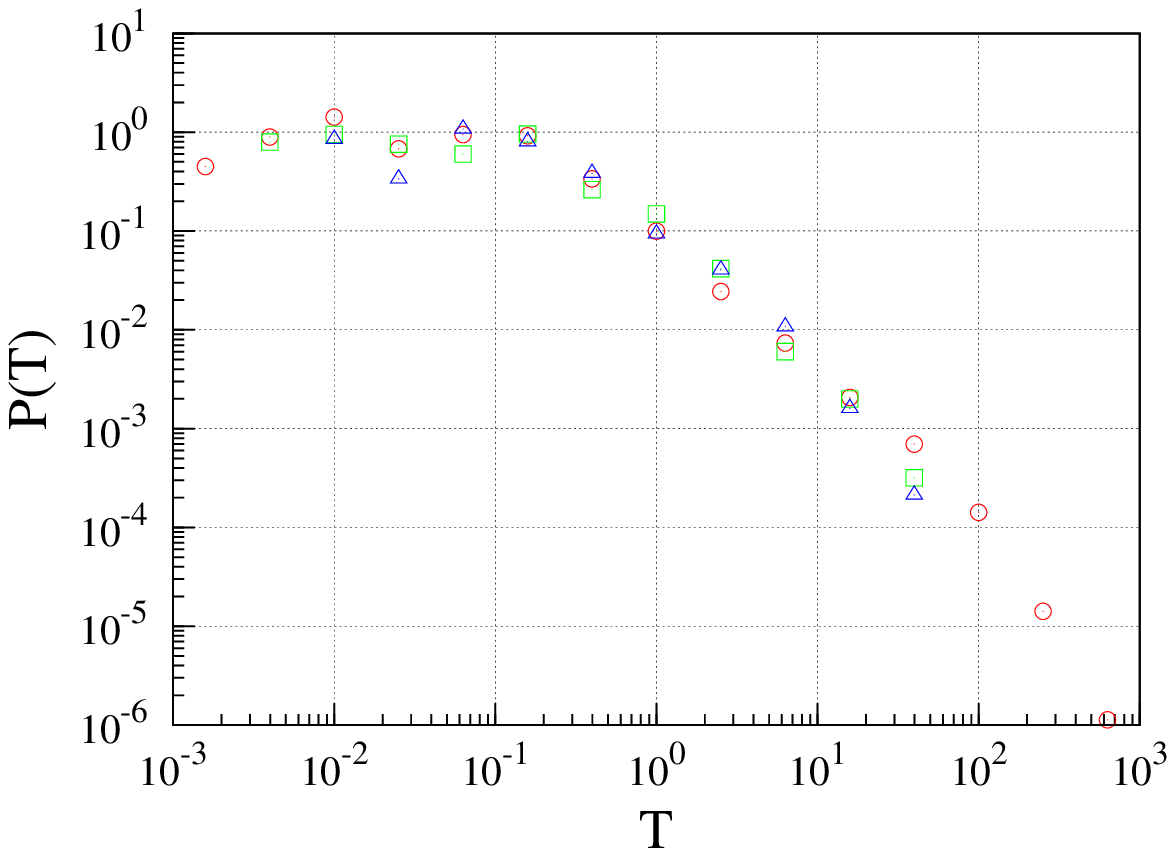}
\includegraphics[width=.4\textwidth]{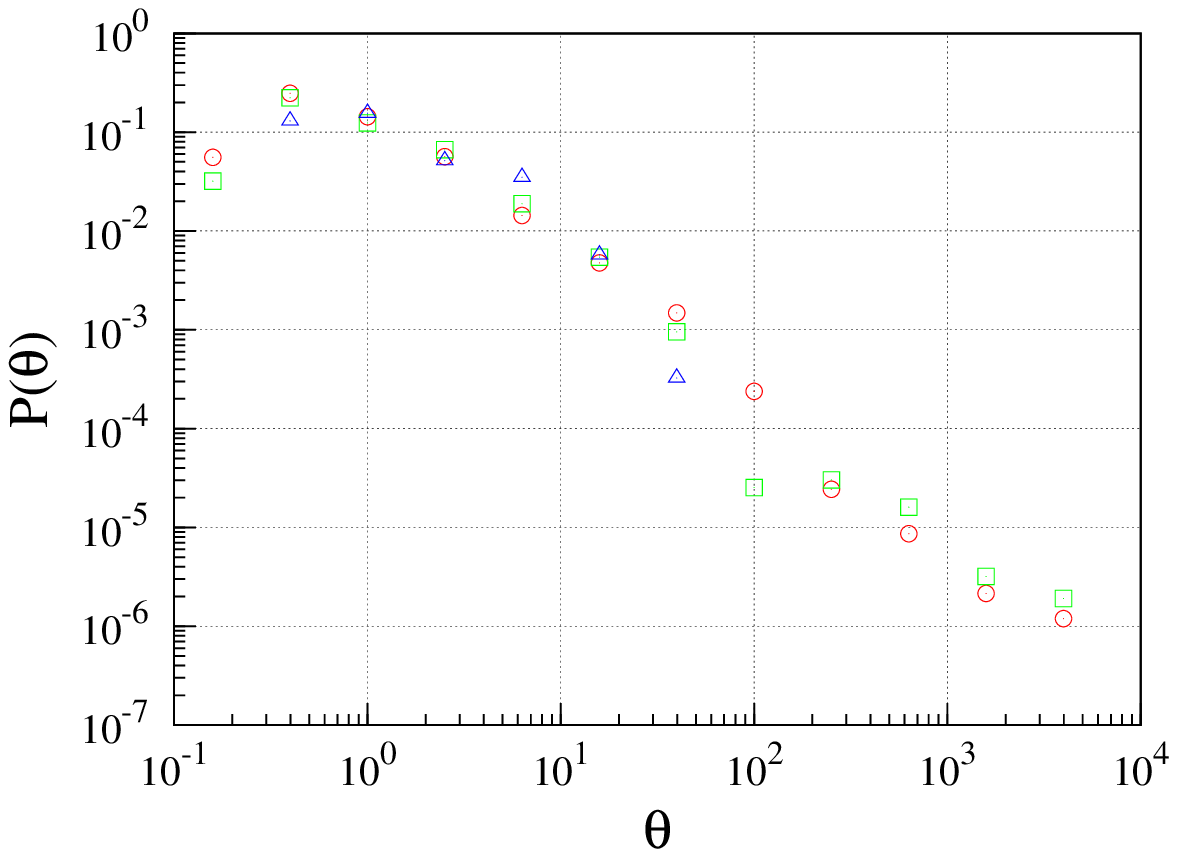}
\caption{Dependence of the burst size $S$ as a function of the burst duration $T$: traces from the top of the figure for size $S$ above the threshold value $x_{th} = 0.02$, 0.1 and 0.5, respectively, distributions of the burst size, $P(S)$, burst duration, $P(T)$, and interburst time, $P(\theta)$, for the peaks above the the threshold value $x_{th} = 0.1$. Calculations are as in figures \ref{figure_2} and \ref{figure_3} with the parameters $\eta =2$, $x_m=10^{-2}$ and different values of the parameter $\lambda$: $\lambda=2$ (circles), $\lambda=3$ (squares) and $\lambda=4$ (triangles). }
\label{figure_4}
\end{figure}

Numerical analysis indicate that the variable of equation \eref{signal} exhibits some peaks, bursts or extreme events, corresponding to the large deviations of the variable from the appropriate average value. 
As examples, in figure \ref{figure_1} we show the illustrations of the signals generated according to equation \eref{signal} for different slopes of the signal distributions and the dependence of the interburst time $\theta_j$ on the burst occurrence number $j$. We see that the the computed signal is composed of bursts of different size with a wide range distribution of the inerburst time. In figures \ref{figure_2} and \ref{figure_3} the numerical calculations of the distribution density, $P\left( x\right)$, power spectral density, $S\left(f\right)$, autocorrelation function, $C(s)$, and the second order structural function, $F_2\left( s\right) =\sqrt{F\left( s\right) }$, for solutions of equation \eref{signal} with $\eta =2$, $x_m=10^{-2}$ and different values of the parameter $\lambda$ are presented. We see rather good agreement between the numerical calculations and the analytical results except for the structural function $F_2\left( s\right)$ when $\lambda>3$. Numerical evaluation of the structural function in a case of the steep power-law distribution is problematic, because in the calculation one needs to average (squared) small difference of the rare large fluctuations. 

In figure \ref{figure_4} we demonstrate numerically that the size of the generated bursts $S$ is approximately proportional to the squared burst duration $T$, i.e., $S \propto T^2$, and asymptotically approximately power-law distributions of the burst size, $P(S) \sim S^{-1.3}$, burst duration, $P(T) \sim T^{-1.5}$ and interburst time, $P(\theta) \sim \theta^{-1.5}$, for the peaks above the threshold value $x_{th}$ of the variable $x(t)$. 

It should be noted that the parameter $\eta =2$ yields in equation \eref{time} the additive noise and the linear relaxation of the signal $x=a / \tau$, i.e., the simple (pure) Brownian motion in the actual time of the interevent time with the linear relaxation of the signal. 

\section{Conclusions}
Starting from the multiplicative point process we obtain the stochastic nonlinear differential equations, which generate signals with the power-law statistics, including $1/f^\beta$ fluctuations. We derive analytical expressions for the probability density of the signal, for the power spectral density, the autocorrelation function, the second oder structural function and demonstrate that the analytical results are in agreement with the results of numerical simulations. The numerical analysis of the equations reveals the secondary structure of the signal composed of peaks or bursts, corresponding to the large deviations of the variable $x$ from the proper average fluctuations. The burst sizes are approximately proportional to the squared duration of the burst. According to the theory \cite{Kuntz,Ruseckas} such dependence for the uncorrelated bursts should result in $1/f^\beta$ noise with $\beta\approx 2$ in the relatively high-frequency region. The power-law distribution $P(\theta)$ of the interburst time $\theta$ indicates in correlation of the burst occurrence times and may result in $1/f^\beta$ noise with $\beta < 2$, similarly to the point process model \cite{Kaulakys,Kaulakys1,Kaulakys2}. On the other hand, the proposed model reproduces $1/f$ noise models and the processes not only in SOC and crackling systems but it is related with the clustering Poisson process \cite{Gruneis}, $1/f$ noise due to diffusion of defects or impurity centers in semiconductors \cite{Gruneis2}, $1/f$ noise in nanochannels, single-channel and ion channel currents \cite{Derksen}, etc. Therefore, the presented and analyzed model may be used for simulation the long-range scaled processes exhibiting $1/f$ noise, power-law distributions and self-organization.

\end{document}